\documentclass{pasj00}

\begin{document}
\SetRunningHead{Tanaka et al.}{On the Rebrightenings of Classical Novae during the Early Phase}
\Received{2010/09/02}
\Accepted{2010/10/27}

\title{On the Rebrightenings of Classical Novae during the Early Phase}

\author{Jumpei \textsc{Tanaka}$^1$, Daisaku \textsc{Nogami}$^2$, Mitsugu \textsc{Fujii}$^3$, Kazuya \textsc{Ayani}$^4$ and Taichi \textsc{Kato}$^1$}
\affil{$^1$Dept. of Astronomy, Kyoto University, Sakyo-ku, Kyoto 606-8502}
\affil{$^2$Kwasan Observatory, Kyoto University, Yamashina-ku, Kyoto 607-8471}
\affil{$^3$Fujii-Bisei Observatory, 4500 Kurosaki, Tamashima, Kurashiki, Okayama 713-8126}
\affil{$^4$Bisei Astronomical Observatory, 1723-70 Okura, Bisei-cho, Ibara, Okayama 714-1411}

\KeyWords{stars: novae, cataclysmic variables} 

\maketitle

\begin{abstract}
We report on the spectral evolution of 6 classical novae, V1186~Sco, V2540~Oph, V4745~Sgr, V5113~Sgr, V458~Vul, and V378~Ser, based on the low-resolution spectra obtained at the Fujii-Bisei Observatory and the Bisei  Astronomical Observatory, Japan. In the light curves, these 6 novae show several rebrightenings during the early phase lasting $\sim$10 days after the first maximum in fast novae, and $\sim$100 days in slow novae. The early spectra of all of these novae had emission lines with a P-Cygni profile at the maximum brightness. The absorption component of the P-Cygni profiles then disappeared after the maximum, and reappeared when the novae brightened to the next maximum. We suggest that the re-appearance of the absorption component at the rebrightening is attributable to re-expansion of the photosphere after it once shifts sufficiently inside. From the light curves, we found that the time intervals of the rebrightenings of these 6 novae show a similar systematic trend, which is applicable to all types of novae: fast and slow, and Fe~II type and hybrid type. Moreover, we note the difference between the spectra at the rebrightenings during the early phase and at the rebrightening in V2362 Cyg, and at the oscillation during the transition phase in V1494 Aql, which means difference of the physical mechanism of the rebrightening during the early phase and the later oscillations.
\end{abstract}

\section{Introduction}
Novae are a kind of cataclysmic variable stars, i.e. close binary systems consisting of a white dwarf and a low-mass normal star (for a review, see \cite{Hellier2001}). The cause of the nova eruption is considered to be the thermonuclear runaway reaction on the surface of the white dwarf. Matters transfered from the secondary are accumulated and compressed on the white dwarf. When the temperature and density increase sufficiently, nuclear reactions occur to cause the nova eruption(\cite{Starrfield1976}).

The nova eruption is a very exciting phenomenon, and the nova systems are the best classically known transient objects.  
Payne-Gaposchkin (1964) reported the classification of the nova light curve based on the time in days for the decline by 2 magnitudes from the maximum, $t_2$.
Duerbeck (1981) reported that there are some light curve types of novae. They suggested a classification system for nova light curves, including dust dips. 
Williams (1992) divided the early post-outburst spectra of novae into two classes, the Fe~II type nova and He/N type nova, based on the emission lines. They proposed that Fe~II type spectrum was formed in a continuous wind, while He/N type spectrum was formed in a discrete shell. 
The diversity of light curves of the novae (see \cite{Kiyota2004},\cite{Strope2010}), however, has not been fully understood.
The initial decline after the maximum is usually smooth in the optical, especially in fast novae. Some slow or moderately fast, but not fast novae, however, have been known to show oscillations having a large amplitude during the early phase (for example HR~Del: \cite{Duerbeck1981}, DO~Aql: \cite{Vogt1928}, RR~Pic: \cite{Spencer1931}), although it is not clear how these novae accomplish these rebrightenings (\cite{Duerbeck1981}). There are, on the other hand, some novae having oscillations during the transition phase. While all of such novae having oscillations during the transition phase have been slow nova (e.g. V4745~Sgr, \cite{Ak2005}), some fast novae showing several peaks during the early phase have been recently discovered (e.g. V458~Vul, \cite{Tarasova2007}). Strope et al.~(2010) reported the  classification and properties of the nova light curve. They reported that there are at least 14 novae, showing the rebrightenings during the early phase, in 93 novae of their sample. The mechanism of these rebrightenings has been still unclear.

In this paper, we present the results of the spectral monitoring of the 6 novae: V1186~Sco, V2540~Oph, V5113~Sgr, V4745~Sgr, V458~Vul, V378~Ser, showing several brightness maximum during the early phase. The details of our targets of our observations are presented in section~2. Our spectral observations are summarized in section~3. The results of the spectral monitoring of the 6 novae are described in section~4. We discuss the properties in the light curves, the spectral evolutions during several rebrightenings, and some interpretations in section~5. Conclusions are stated in section~6.

\section{On the targets}
In this section we briefly summarize the research history of our targets.
\subsection{V1186 Sco}
V1186~Sco was discovered at 2004 July 3.1 (UT) by Pojmanski et al.~(2004) at $\mathrm{V} = 11.98$. We then obtained the earliest spectrum at 2004 July 6.51 (UT), about 3 days before the visual maximum. This spectrum is presented later in this paper. Schwarz et al.~(2007) reported Target of Opportunity observations by Spitzer Space Telescope. They also showed that $t_2$, the time to decline by
 2~mag, was 20~days. They derived an absolute magnitude $M_\mathrm{V} = -5.5 \pm 0.5$, using the best-fit photoionization model. They reported that $\mathrm{E(B-V)} = 0.45 \pm 0.1$ and the derived distance was $5.5 \pm 0.5$ kpc. Burlak \& Henden~(2008) derived $t_2$ of 17 days and $t_3$ of 46 days, suggesting that V1186 Sco is a fast nova. They also estimated the absolute magnitude $M_\mathrm{V} = -7.8$.

\subsection{V2540 Oph}
V2540~Oph was independently discovered by Katsumi Haseda and Yuji Nakamura on 2002 January 24 at $\mathrm{V} = 9.0$ (see \cite{Nakamura2002}). Seki et al.~(2002) reported that this nova already reached $\mathrm{V} = 8.9$ on 2002 January 19. Retter et al.~(2002) classified this nova as an Fe~II type nova, based on H and Fe~II lines. Kato et al.~(2002) reported that the absolute magnitude of the nova progenitor was $M_\mathrm{V} = 5.7$ and the expected absolute V-band magnitude at the maximum was $M_\mathrm{V} = -7.0 \pm 0.5$. In addition, they proposed that the nova should have either a short orbital period or a high inclination. Ak et al.~(2005) estimated the maximum absolute visual magnitude of $M_\mathrm{V} = -6.2 \pm 0.4$ mag and the distance of $5.2 \pm 0.8$ kpc. Moreover, they derived its orbital period of $0.284781 \pm 0.000006$ d, and the secondary mass of about $0.75 \pm 0.04 {M_\odot}$ from the mass-period relation of Smith \& Dhillon (1998). Burlak \& Henden~(2008) derived $t_2$ of 167 days and $t_3$ of 305 days, suggesting that V2540 Oph is a slow nova. They also estmiated the absolute magnitude $M_\mathrm{V} = -5.5$.

\subsection{V5113 Sgr}
V5113~Sgr was discovered by Brown et al.~(2003) at 2003 September 17.52 (UT) at 9.2~mag (\cite{Brown2003b}). They estimated that this nova was 8.8~mag at September 18.43. They also reported that there were H$\alpha$ lines in emission with a narrow absorption component of the P-Cygni profile blue-shifted by 590 km s$^{-1}$ at September 19.104 (UT). Ruch et al.~(2003) reported that there were H$\beta$, H$\gamma$, H$\delta$ and Fe~II lines with the absorption components of P-Cygni profiles at September 19.10 (UT). The thermal velocity of the absorption component was 800 km s$^{-1}$. The optical and IR spectra obtained 9 months after outburst contained the coronal lines of [Si~VI] and [S~VIII] as well as He~II lines (\cite{Rudy2004}).

\subsection{V4745 Sgr}
V4745~Sgr was independently discovered by Brown and Yamamoto on 2003 April 25 and 26 (see \cite{Brown2003a}). Ashock \& Banerjee (2003) reported the spectral observation showing Balmer series with P-Cygni profiles and classified this nova as an Fe~II type nova. Cs$\acute{\rm a}$k et al. (2005) reported that P-Cygni profiles appeared at and around peaks of rebrightenings. They suggested that this behavior was due to multiple episodes of mass ejection. They derived the absolute magnitude, $-8.3 \pm 0.5$ mag and the distance of 9 kpc$<d<$19 kpc. Dobrotka et al.~(2006) reported that the orbital period is $0.20782 \pm 0.000003$ d, above the period gap, and the beat period between the orbital and spin period is $0.017238 \pm 0.000037$ d. They suggested that this nova was an intermediate polar candidate, and obtained the secondary mass of $0.52 \pm 0.05 M_{\odot}$.

\subsection{V458 Vul}
V458~Vul was discovered by H.Abe at 2007 August 8.54 (UT) (see \cite{Nakano2007}). The spectra were acquired by Fujii at 2007 August 9.48 (UT) (see \cite{Buil2007}), which is presented later in this paper. Poggiani (2008) reported that V458 Vul is a fast nova, with $t_2$ and $t_3$ of 7 and 15 days, respectively. They derived the absolute magnitude of $M_\mathrm{V} = -8.8$, and a roughly estimated distance of 6.7-10.3 kpc. They classified this nova as a hybrid nova, which evolved from the Fe~II nova toward the He/N nova. Arai et al.~(2009) reported the spectral evolution of V458 Vul. They found that the absorption component of the P-Cygni profile was present at maximum brightness of the rebrightening. Moreover, they reported that He~I lines showed  unnatural double-peaks during the early phase. Rodr$\acute{\rm \i}$guez-Gil et al.~(2010) reported the optical spectroscopy during 15 months starting 301 days after the discovery. They derived the orbital period of 98.09647 $\pm$ 0.00025 min by using the radial velocity curves of He~II lines. They proposed that V458~Vul is the planetary nebula central binary star with the shortest period known. They suggested that V458~Vul is a post-double common-envelope binary system composed of a $M_1 \geq 1.0 M_{\odot}$ white dwarf and a $M_2 \sim 0.6 M_{\odot}$, post-AGB star. 

\subsection{V378 Ser}
V378~Ser was discovered by Pojmanski at 2005 March 14.389 (UT) (\cite{Pojmanski2005}). Ederoclite et al.~(2005) reported the spectroscopic observations carried out at April 5.38 (UT) at La Silla with a 2.2-m telescope. They identified this nova as an Fe~II type nova, with a spectrum dominated by strong H$\alpha$ and O~I~7773 and 8446. In addition, they reported that there were Fe~II, Na~I and Ca I lines in emission. All the Balmer and O~I lines were flanked by strong, double absorption components of the P-Cygni profiles.

\section{Observation}
Most of our low dispersion spectra with a resolution of $\lambda / \Delta \lambda \approx 600$ (at 5852~\AA) have been taken with the FBSPEC1 and FBSPEC2, which have been developed by one of the authors (MF),  attached to the 28 cm telescope of the Fujii-Bisei Observatory (Okayama, Japan). The spectrum of V2540 Oph on 2002 January 27 was taken with a low-resolution spectrograph ($\lambda/\Delta\lambda\approx 1,000$), attached to the 101 cm telescope of the Bisei Astronomical  Observatory (Okayama, Japan). The spectral coverage is 3800 to 8400 \AA. The spectra of each object were obtained in 2 - 10 nights during the early phase when the targets were brighter than V = 10-11 mag. The signal-to-noise ratio is typically 10 to 30. The reducton of the spectra was carried out with NOAO~IRAF. Table \ref{tab_observation} gives the journal of our specroscopic observations.

\begin{longtable}{llll}
  \caption{Spectroscopic Observations of Classical Novae Showing Several Peaks}\label{tab_observation}
  \hline              
   Object name & Date & Spectral range & JD \\
               &      &  (\AA) & (2450000$+$) \\
\endfirsthead
  \hline
\endhead
  \hline
\endfoot
  \hline
\endlastfoot
  \hline
  V1186~Sco & 2004 July 6 & 3800-8300 & 3193.10 \\
  V1186~Sco & 2004 July 13 & 3900-8400 & 3200.01 \\
  V1186~Sco & 2004 July 14 & 3900-8400 & 3201.00 \\
  V1186~Sco & 2004 July 20 & 3900-8400 & 3206.99 \\
  V1186~Sco & 2004 July 26 & 3900-8400 & 3213.00 \\
  V1186~Sco & 2004 August 5 & 3900-8400 & 3223.02 \\
  V2540~Oph & 2002 January 27 & 4800-6800 & 2302.40 \\
  V2540~Oph & 2002 February 24 & 4500-7100 & 2330.31 \\
  V2540~Oph & 2002 April 12 & 4400-7100 & 2377.26 \\
  V2540~Oph & 2002 May 1 & 4400-7100 & 2396.12 \\
  V2540~Oph & 2002 May 19 & 4400-7100 & 2414.14 \\
  V5113~Sgr & 2003 September 21 & 3900-8400 & 2903.96 \\
  V5113~Sgr & 2003 September 22 & 3900-8400 & 2904.96 \\
  V5113~Sgr & 2003 September 27 & 3900-8400 & 2909.93 \\
  V5113~Sgr & 2003 September 30 & 3900-8500 & 2912.95 \\
  V5113~Sgr & 2003 October 8 & 3900-8500 & 2920.90 \\
  V5113~Sgr & 2003 October 24 & 3900-8400 & 2936.88 \\
  V4745~Sgr & 2003 April 28 & 4600-7200 & 2758.26 \\
  V4745~Sgr & 2003 April 30 & 4600-7200 & 2760.28 \\
  V4745~Sgr & 2003 May 8 & 3900-8400 & 2768.29 \\
  V4745~Sgr & 2003 May 28 & 3900-8400 & 2788.23 \\
  V458~Vul & 2007 August 9 & 3800-8300 & 4321.98 \\
  V458~Vul & 2007 August 10 & 3800-8300 & 4323.15 \\
  V458~Vul & 2007 August 12 & 3800-8200 & 4325.16 \\
  V458~Vul & 2007 August 15 & 3800-8200 & 4328.03 \\
  V458~Vul & 2007 August 16 & 3800-8200 & 4329.03 \\
  V458~Vul & 2007 August 25 & 3800-8300 & 4338.01 \\
  V458~Vul & 2007 September 5 & 3800-8300 & 4349.12 \\
  V458~Vul & 2007 September 10 & 3800-8200 & 4353.99 \\
  V458~Vul & 2007 September 25& 4600-6700 & 4368.91 \\
  V458~Vul &2007 October 6 & 4600-6700 & 4379.99 \\
  V378~Ser & 2005 April 4 & 3900-8400 & 3465.30 \\
  V378~Ser & 2005 April 22 & 3900-8400 & 3483.24 \\
\end{longtable}

\section{Result}
\subsection{V1186~Sco}
\begin{figure}
  \begin{center}
    \FigureFile(80mm,50mm){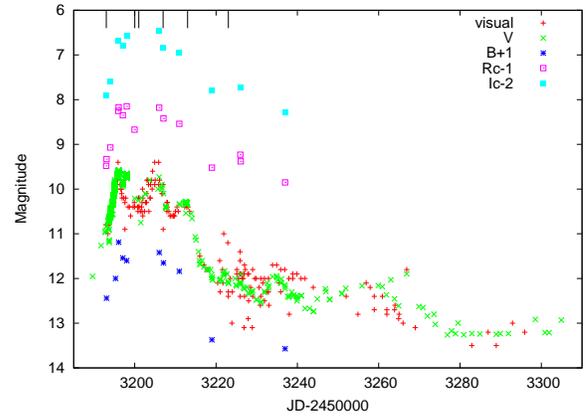}
  \end{center}
  \caption{Light curve of V1186~Sco. The photometry was taken from VSNET, AAVSO and ASAS-3. The epochs of our spectroscopic observations are marked with the tick marks.}\label{v11861}
\end{figure}

\begin{figure*}
  \begin{center}
    \FigureFile(160mm,100mm){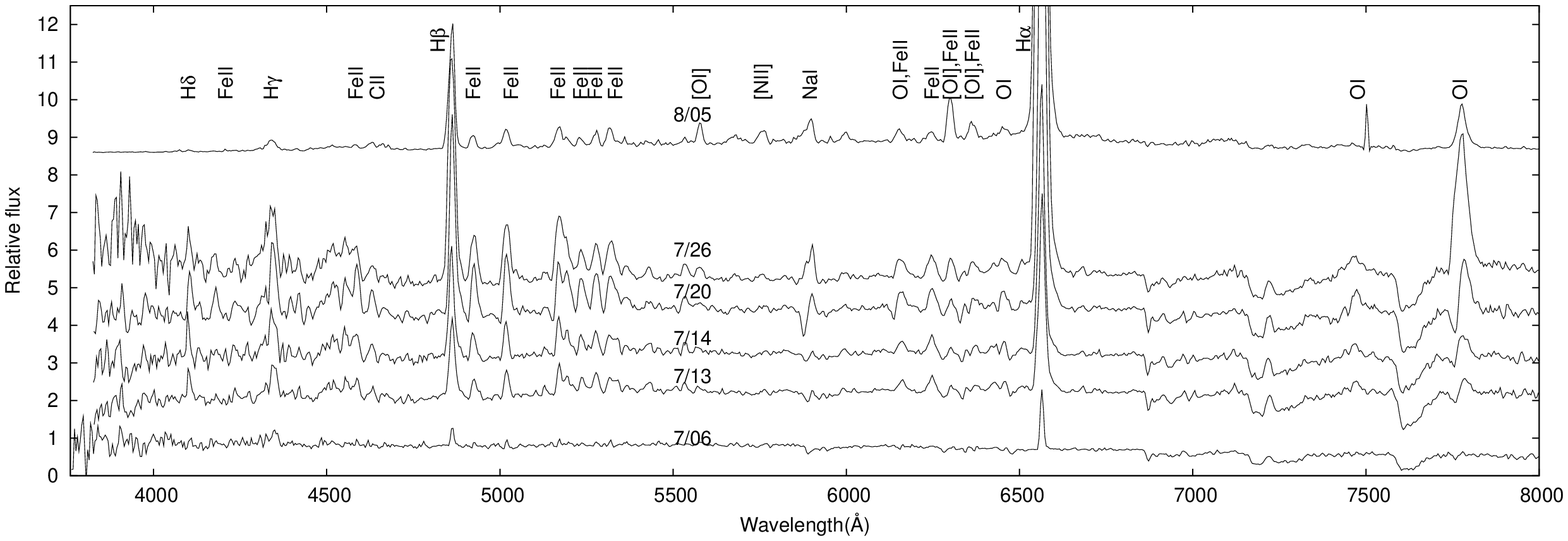}
  \end{center}
  \caption{Normalized spectra of V1186 Sco from 2004 July 6 to August 5. For visuality, the data after July 6 are vertically shifted.}\label{v11862}
\end{figure*}

Figure \ref{v11861} shows the light curve of V1186~Sco collected from the VSNET (\cite{Kato2004}), AAVSO\footnote{http://www.aavso.org} and ASAS-3 (\cite{Pojmanski2002}) archives. This nova shows three peaks during the early phase.
The spectra were obtained in 6 nights between July 6 (JD 2453193.10) and August 5 (JD 2453223.02), 2004. All of these spectra are presented in figure \ref{v11862}. On July 6 (JD 2453193.10), before the first maximum brightness, the spectrum covering 3800 \AA$\;$ to 8400 \AA$\;$ is characterized by strong emission lines of the Balmer series. The next spectra were obtained on July 13 and 14 (JD 2453200.01 and 2453201.00), after the first maximum brightness. Fe~II and O~I lines appear in emission. Thus this nova can be classified as an Fe~II class object. On July 20 (JD 2453206.99), just at the second maximum brightness, we identified  Na~I~D line accompanied by a P-Cygni profile having a velocity of about $-$1250 km s$^{-1}$ between the peaks of the absorption and emission components. This absorption feature is also present in Fe~II emission lines. On July 26 (JD 2453213.00), at the third maximum, we can identify the Balmer series and Na~I, Fe~II and O~I lines. The weak absorption component of the P-Cygni profile is present in Fe~II, and Na~I~D line. 
O~I~7773 has developed. The last spectrum on August 5 (JD 2453223.02) shows the presence of the  emission lines of H, Na~I, Fe~II, O~I, [N~II] and [O~I]. O~I and [O~I] lines are stronger compared to the other lines.

\subsection{V2540~Oph}
\begin{figure}
  \begin{center}
    \FigureFile(80mm,50mm){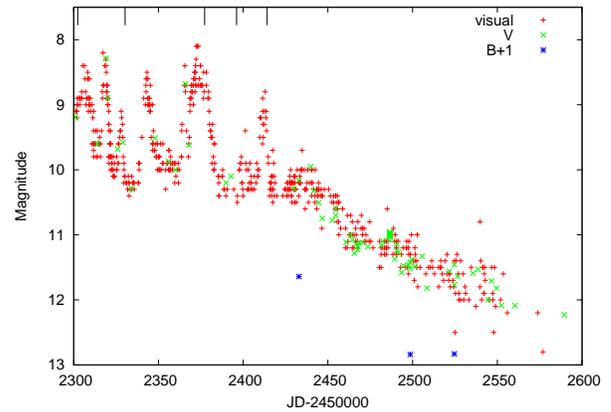}
  \end{center}
  \caption{Light curve of V2540~Oph taken from VSNET, AAVSO and ASAS-3. The epochs of our spectroscopic observations are marked with the tick marks.}\label{v25401}
\end{figure}

\begin{figure*}
  \begin{center}
    \FigureFile(160mm,100mm){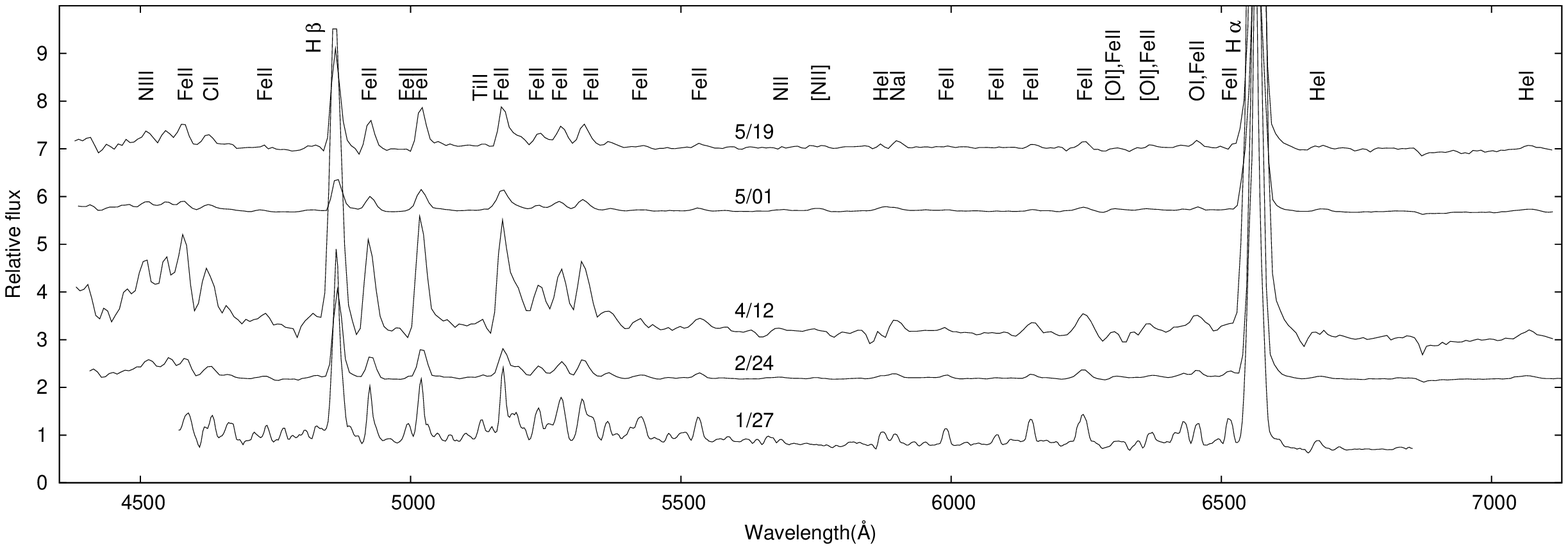}
  \end{center}
  \caption{Normalized spectra V2540 Oph from 2002 January 27 to May 19. For visuality, the data after January 27 are vertically shifted.}\label{v25402}
\end{figure*}

\begin{figure}
  \begin{center}
    \FigureFile(80mm,50mm){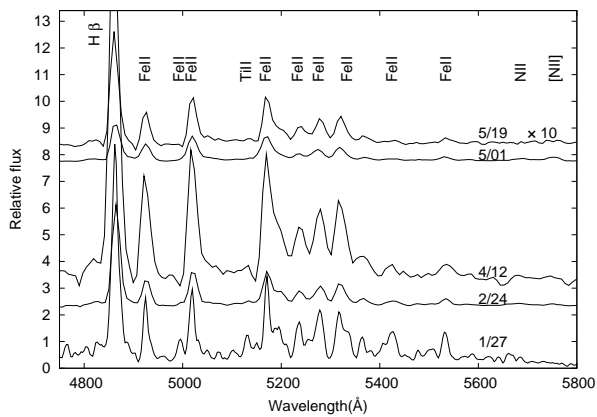}
  \end{center}
  \caption{Normalized spectra of V2540 Oph from 2002 January 27 to May 19. The spectra are the same as those in figure 4, but the part around 5000$\AA$ is expanded. We can see reappearance of the absorption component of Fe II lines at the 6th rebrightening on 2002 May 19.}\label{v25403}
\end{figure}

Figure \ref{v25401} shows the light curve of V2540~Oph. This nova exhibits several brightness maxima  during the early phase. We obtained the spectra of V2540~Oph in 5 nights between January 27 (JD 2452302.40) and May 19 (JD 2452414.14), 2002. All of the spectra are displayed in figure \ref{v25402}. The first observation was carried out on January 27 (JD 2452302.40), at the pre-maximum. We can identify the emission lines of H, Fe~II, O~I, [O~I], Na~I, He~I, N~III, C~II and Ti~II. On February 24 (JD 2452330.31), just after the third brightness maximum, the spectrum shows the presence of emission lines of H, Fe~II, [Fe~II], O~I, [O~I], Na~I, He~I, N~III, and C~II. The spectrum suggests that this nova is classified as an Fe~II nova. On April 12 (JD 2452377.26), on the fifth brightness maximum, we identified the same emission lines as the previous spectrum. The profiles of Fe~II lines, however, show absorption components of the P-Cygni profiles at about $-$1310 km s$^{-1}$ and about $-$1230 km s$^{-1}$ from the peak of the emission component for Fe~II~5018, and 5169 lines, respectively. These P-Cygni profiles disappear at the next observation on May 1, between the fifth maximum of brightness and the sixth brightness maximum, and then reappear at about $-$1300 km s$^{-1}$ and about $-$1090 km s$^{-1}$ for Fe~II~5018, and 5169 lines, respectively on May 19 (JD 2452414.14), just at the sixth brightness maximum (figure \ref{v25403}). The last spectrum, observed on May 19 (JD 2452414.14), hardly exhibits the other changes, compared with the last observation.

\subsection{V5113~Sgr}
\begin{figure}
  \begin{center}
    \FigureFile(80mm,50mm){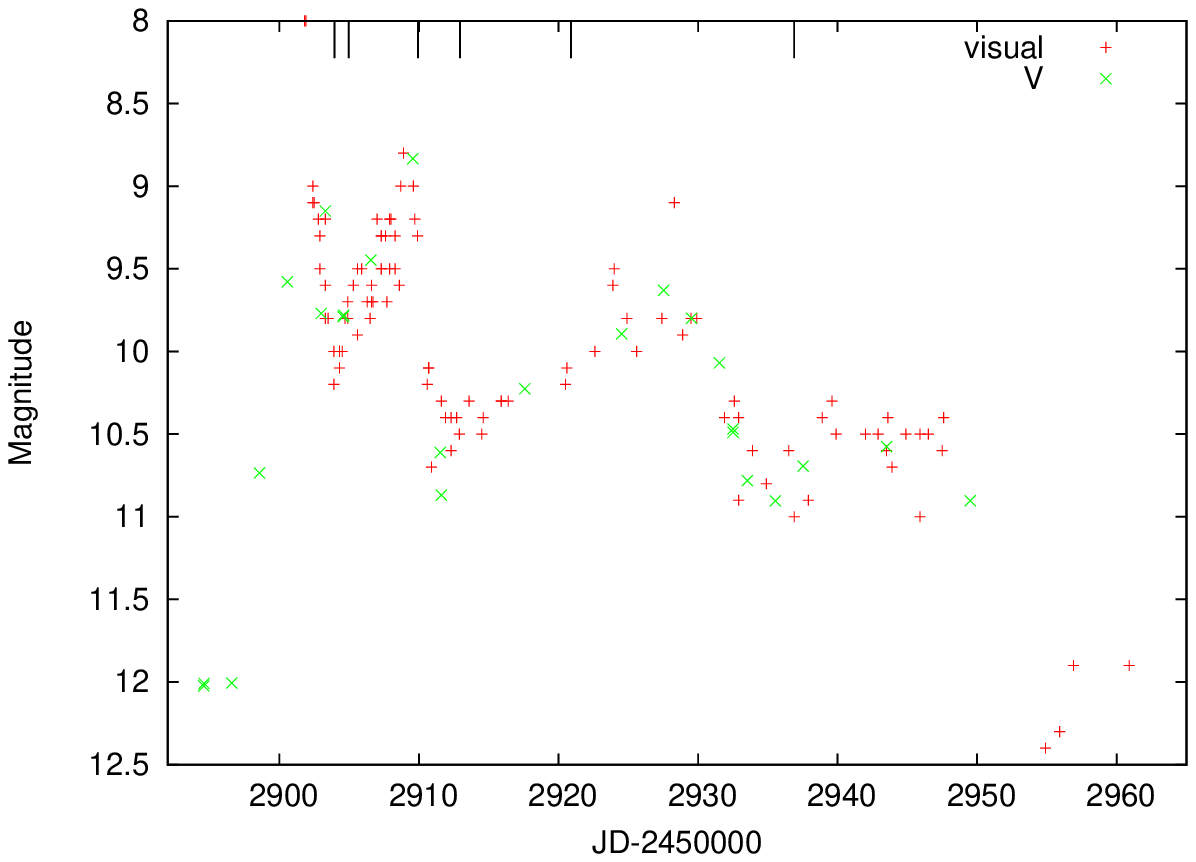}
  \end{center}
  \caption{Light curve of V5113~Sgr taken from VSNET, AAVSO and ASAS-3. The epochs of our spectroscopic observations are marked with the tick marks.}\label{v51131}
\end{figure}

\begin{figure*}
  \begin{center}
    \FigureFile(160mm,100mm){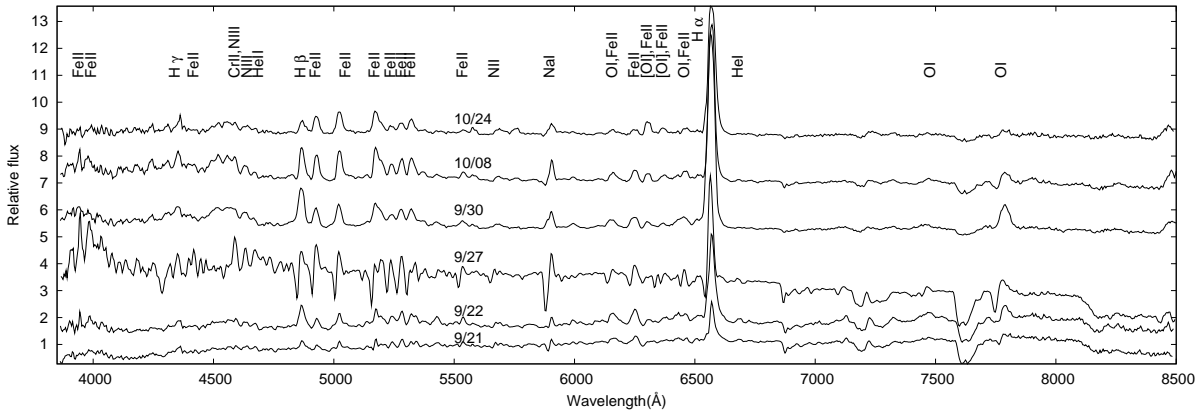}
  \end{center}
  \caption{Normalized spectra of V5113 Sgr from 2003 September 21 to October 24. For visuality, the data after September 21 are vertically shifted.}\label{v51132}
\end{figure*}

Figure \ref{v51131} displays the light curve of V5113~Sgr. This nova also exhibits several peaks.
The spectra were obtained on 6 nights between September 21 (JD 2452903.96) and October 24 (JD 2452936.88), 2003. On September 21 (JD 2452903.96), after the first brightness maximum, the spectrum shows the features of Fe~II type novae, having hydrogen Balmer series and Fe~II and O~I lines with P-Cygni profiles (figure \ref{v51132}). We also identified Na~I~D and N~II lines. The velocity measured from the peaks of the absorption and emission components is about $-$830 km s$^{-1}$ and about $-$680 km s$^{-1}$ for Fe~II~5169, and Na~I~D, respectively. On September 22 (JD 2452904.96), the spectrum hardly changed, but absorption components of the P-Cygni profiles became weaker. On September 27 (JD 2452909.93), at the second brightness maximum, the spectrum changed considerably and showed many lines consisting of Balmer series, Fe~II, O~I and N~II lines with strong absorption components of the P-Cygni profiles. The blue-shifted velocities of the P-Cygni profiles of Na~I~D became higher, about $-$1240 km s$^{-1}$ , while that of Fe~II~5169 was about $-$750 km s$^{-1}$, virtually the same as the velocity on September 21 (JD 2452903.96). The spectrum on September 30 (JD 2452912.95) exhibits the presence of the same emission lines as in previous spectra, while the absorption components of the P-Cygni profiles are weaker. On October 8 (JD 2452920.90), before the third brightness maximum, the spectrum exhibits the emission lines of H, Fe~II, O~I, Na~I, N~II, N~III, [O~I], [O~II] and [N~II]. In addition, absorption components of the P-Cygni profiles reappear in Fe~II and Na~I lines at about $-$680 km s$^{-1}$ and about $-$1280 km s$^{-1}$ respectively. On October 24 (JD 2452936.88), after the third brightness maximum, C~II emission line (7234 \AA) become stronger, and H$\beta$ weaken compared to Fe~II lines. P-Cygni profiles are still present. Each blue-shift velocity is higher, presented in table \ref{tab_pcygni}.

\subsection{V4745~Sgr}
\begin{figure}
  \begin{center}
    \FigureFile(80mm,50mm){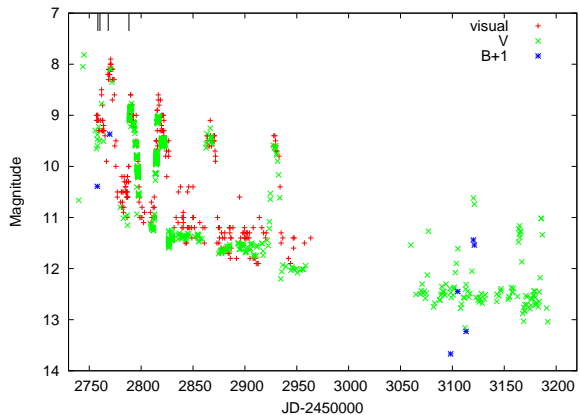}
  \end{center}
  \caption{Light curve of V4745~Sgr taken from VSNET, AAVSO and ASAS-3. The epochs of our spectroscopic observations are marked with the tick marks.}\label{v47451}
\end{figure}

\begin{figure*}
  \begin{center}
    \FigureFile(160mm,100mm){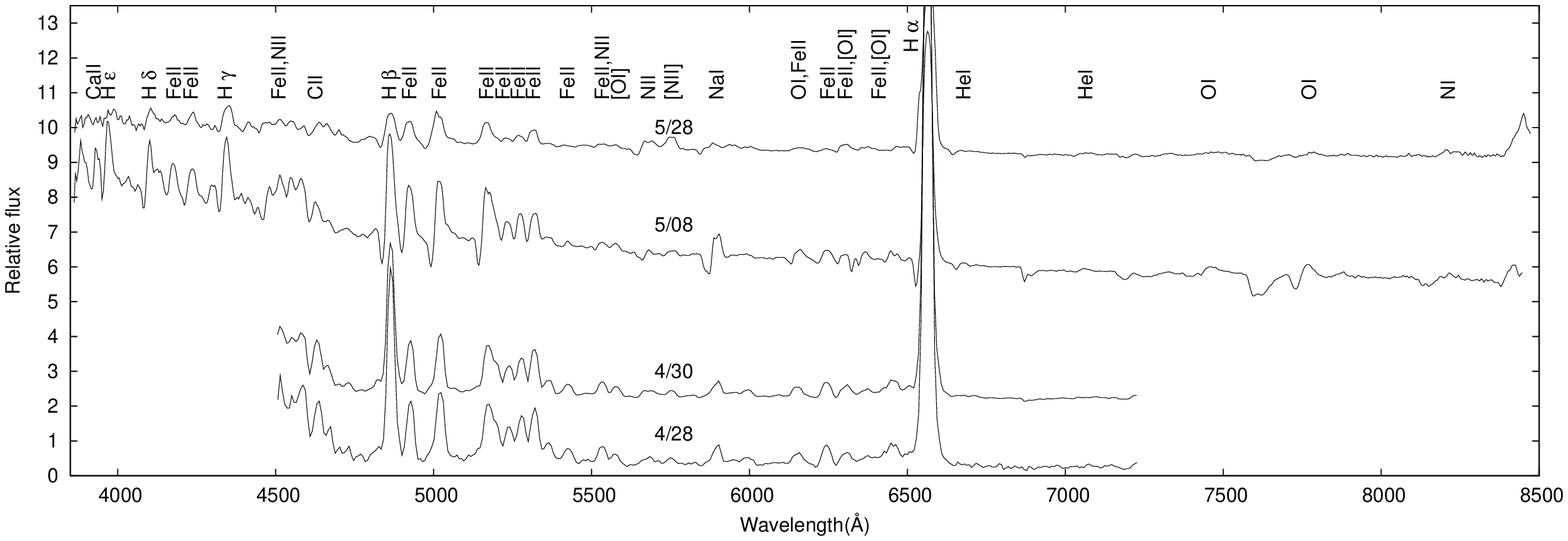}
  \end{center}
  \caption{Normalized spectra of V4745 Sgr from 2003 April 28 to May 28. For visuality, the data after April 28 are vertically shifted.}\label{v47452}
\end{figure*}

Figure \ref{v47451} displays the light curve of V4745~Sgr. This nova shows six (and possibly more) peaks. We obtained the spectra on 4 nights between April 28 (JD 2452758.26) and May 28 (JD 2452788.23), 2003 (figure \ref{v47452}). On April 28 and 30 (JD 2452758.26 and 2452760.28), just after the first brightness maximum, Balmer series and several Fe~II lines appear in emission, indicating that this nova is an Fe~II nova. In addition, there are Na~I, N~II, [N~II], O~I, [O~I] and C~II lines, while, strangely, there are no prominent lines on a redder region than H$\alpha$. On May 8 (JD 2452768.29), at the second brightness maximum, the spectrum has remarkably changed compared to the previous spectrum. Many lines show P-Cygni profiles with a blue-shifted absorption component. The blue-shift velocities measured by using the peaks of the absorption and emission components are about $-$1700 km s$^{-1}$, depending on the lines (table \ref{tab_pcygni}). The exisitence of He~I lines indicates that this nova may be a hybrid nova, not an Fe~II nova. On May 28 (JD 2452788.23), just at the third brightness maximum, the  Balmer lines exhibit P-Cygni profiles, having higher blue-shift velocities. Cs$\acute{\rm a}$k et al. (2005) reported that the absorption component of P-Cygni profiles weakened between two peaks on May 8 (JD 2452768.29) and May 28 (JD 2452788.23). We can thus conclude that P-Cygni profiles once disappeared after the maximum and then reappeared on the next third brightness maximum.

\subsection{Other objects showing several peaks}
\begin{figure}
  \begin{center}
    \FigureFile(80mm,50mm){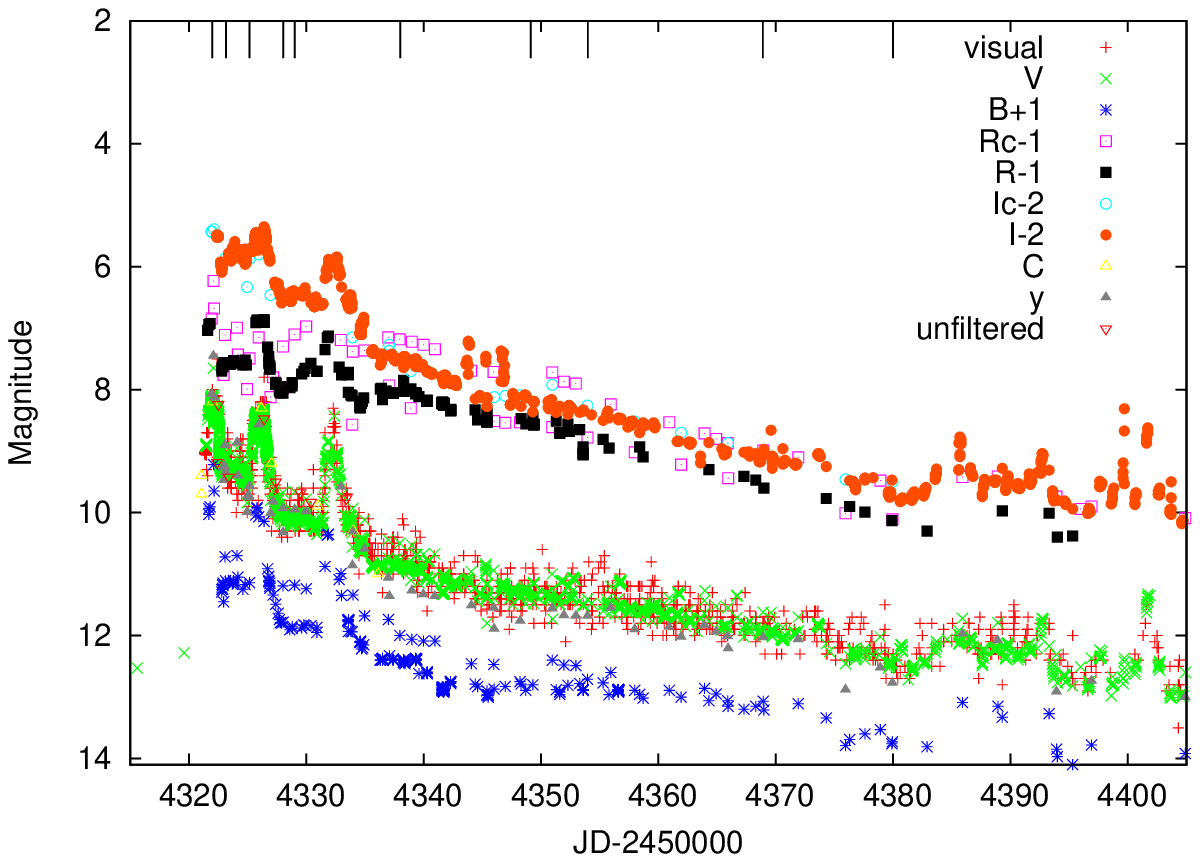}
  \end{center}
  \caption{Light curve of V458~Vul taken from VSNET, AAVSO and ASAS-3. The epochs of our spectroscopic observations are marked with the tick marks.}\label{v4581}
\end{figure}

\begin{figure*}
  \begin{center}
    \FigureFile(160mm,100mm){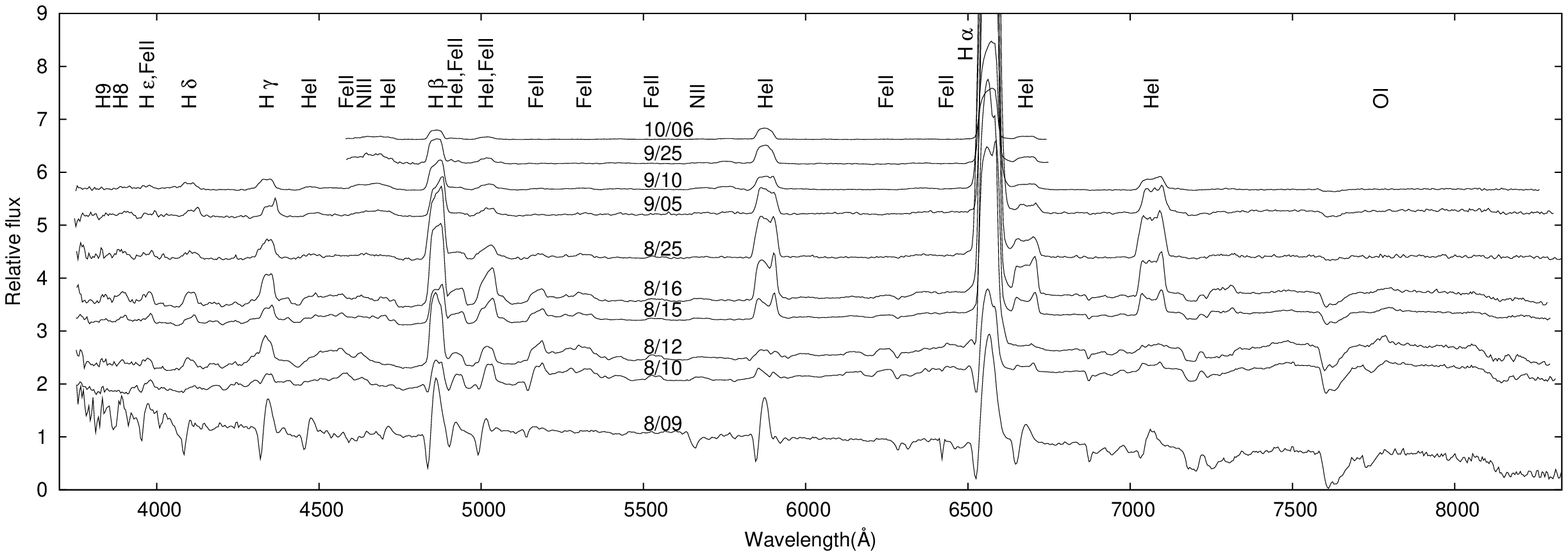}
  \end{center}
  \caption{Normalized spectra of V458 Vul from 2007 August 9 to October 6. For visuality, the data after August 9 are vertically shifted.}\label{v4582}
\end{figure*}

V458~Vul and V378~Ser also show several rebrightenings during the early phase. We obtained spectra of these novae on 10 nights and 2 nights respectively. We did not however take the spectra just on the brightness maxima except the first maximum. The light curve of V458~Vul, presented in figure \ref{v4581}, shows three rebrightenings during the early phase. In addition, this exhibits an oscillation-like feature during the transition phase, which is similar to the behavior of V1494 Aql (see e.g., \cite{Iijima2003}). On August 9 (JD 2454321.98), just at the initial brightness maximum, Balmer series and many He~I lines appear in emission (figure \ref{v4582}). We can identify strong absorption components of the P-Cygni profiles of these lines. On August 10 (JD 2454323.15), there are also Na~I, Fe~II, O~I and N~II lines. Therefore we can classify this nova as a hybrid nova. Here, P-Cygni profiles are not present in our observations except the spectrum on August 9 (JD 2454321.98) and 10 (JD 2454323.15).

\begin{figure}
  \begin{center}
    \FigureFile(80mm,50mm){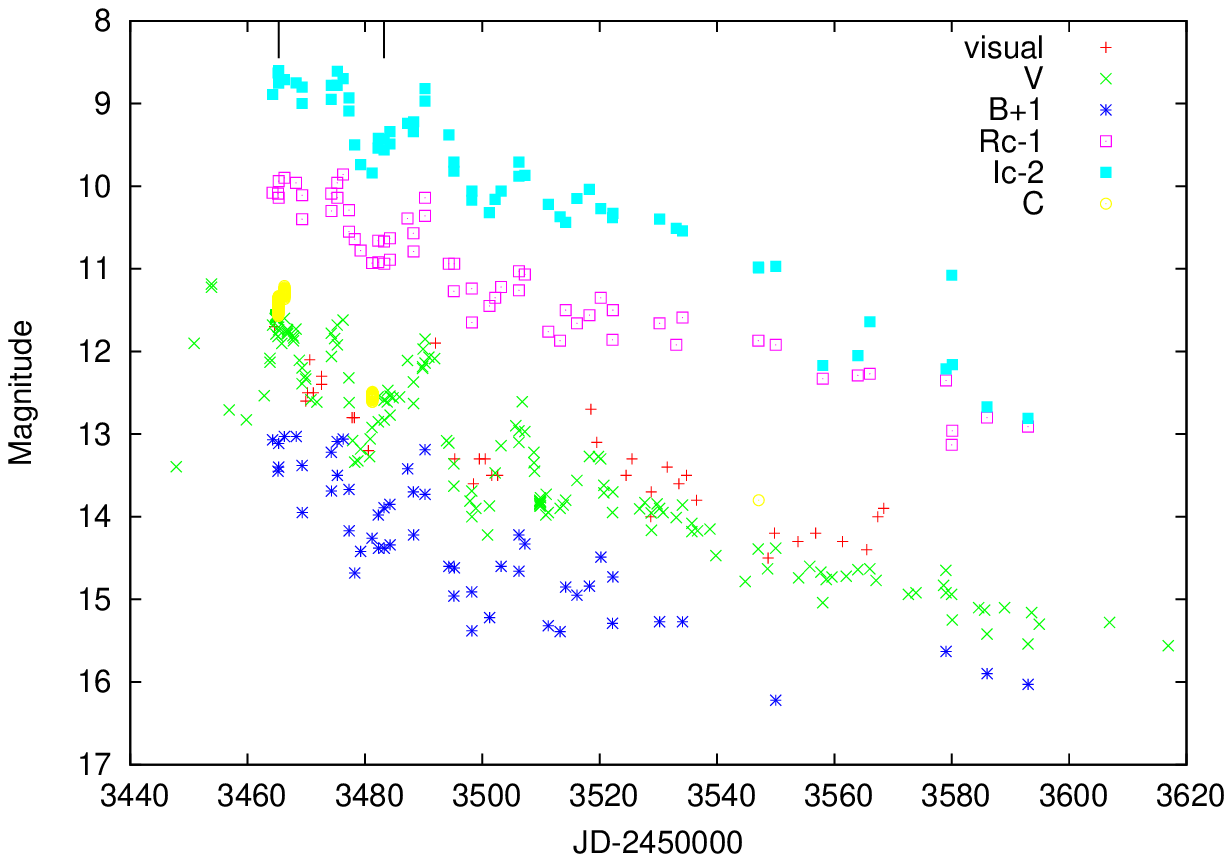}
  \end{center}
  \caption{Light curve of V378~Ser taken from VSNET, AAVSO and ASAS-3. The epochs of our spectroscopic observations are marked with the tick marks.}\label{v3781}
\end{figure}

\begin{figure}
  \begin{center}
    \FigureFile(80mm,50mm){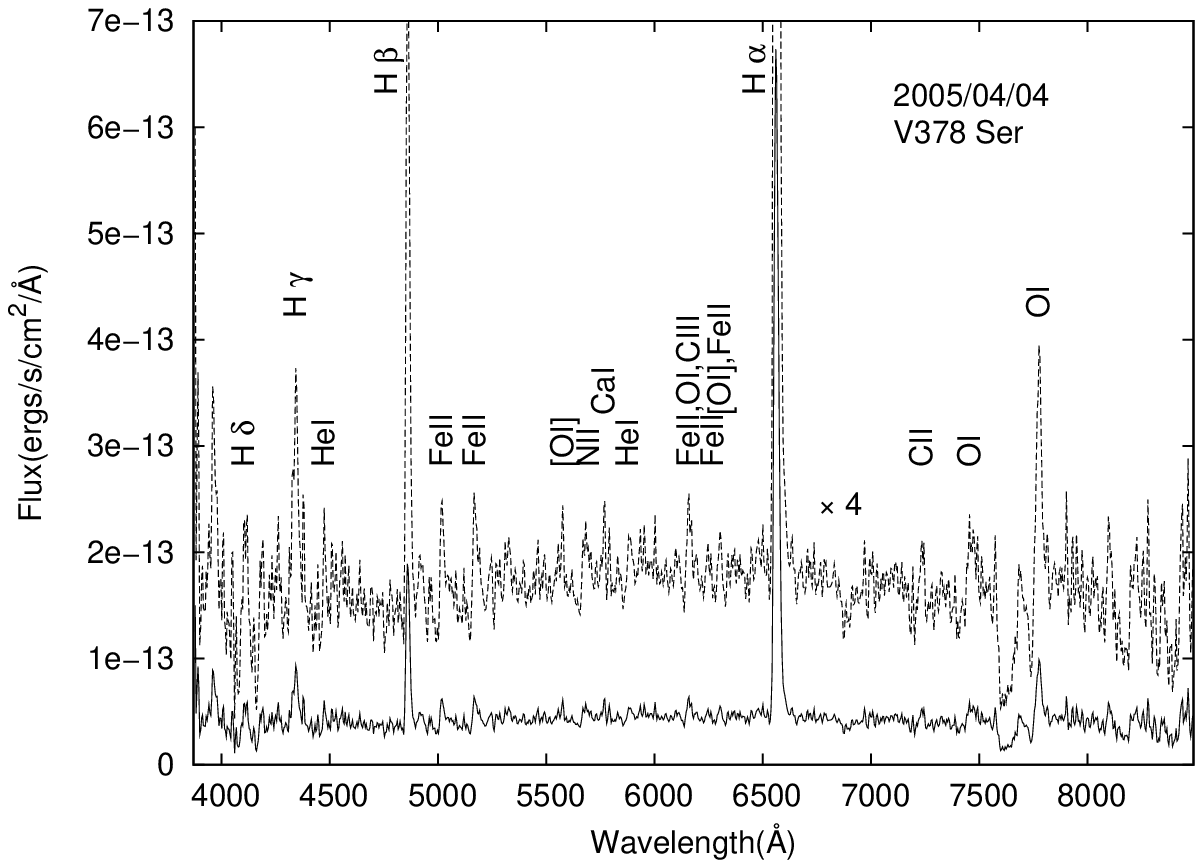}
  \end{center}
  \caption{Spectrum of V378~Ser on 2005 April 4.}\label{v3782}
\end{figure}

\begin{figure}
  \begin{center}
    \FigureFile(80mm,50mm){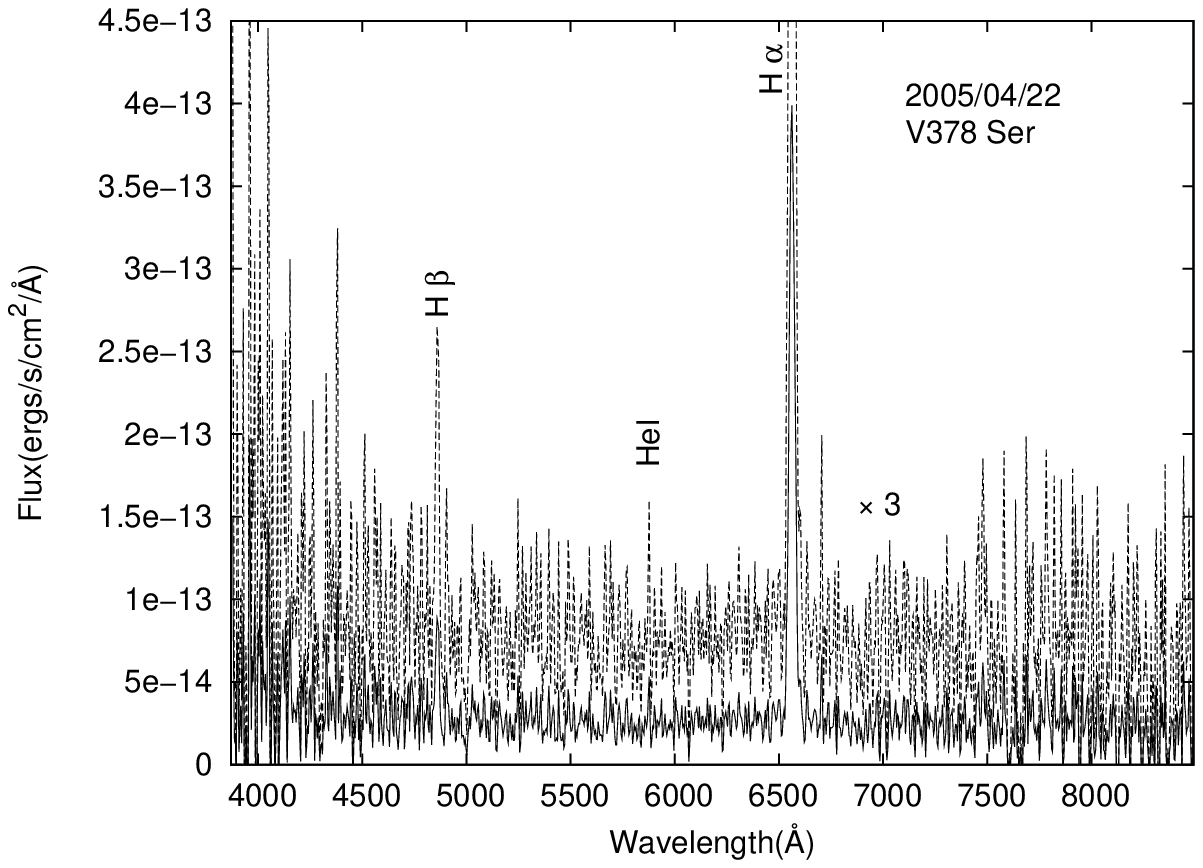}
  \end{center}
  \caption{Spectrum of V378~Ser on 2005 April 22.}\label{v3783}
\end{figure}

The light curve of V378~Ser is presented in figure \ref{v3781}. On April 4 (JD 2453465.30), there are the Balmer series and Fe~II, O~I, N~II, He~I, C~III, [OI] and possibility Ca~I lines (figure \ref{v3782}). The absorption components of P-Cygni profiles appear in H$\alpha$, H$\beta$ and Fe~II lines. On April 22 (JD 2453483.24), it is difficult to discriminate emission lines other than H and He because of the poor signal-to-noise ratio (figure \ref{v3783}).

\begin{longtable}{llllll}
  \caption{Blue-shift velocities (km s$^{-1}$) of absorption components of the P-Cygni profiles}\label{tab_pcygni}
\endfirsthead
 \hline
\endhead
  \hline
\endfoot
  \hline
\endlastfoot
  \hline
  V1186~Sco & & & & & \\
  Atomic identification & Rest wavelength (\AA) & 7/20 & & & \\
  \hline
  Fe~II & 5018 & $-1000 \pm 100$ & & & \\
  Fe~II & 5169 & $-1000 \pm 100$ & & & \\
  Na~I & 5896 & $-1220 \pm 100$ & & & \\
  O~I  & 6155 & $-1070 \pm 100$ & & & \\
  \hline

  & & & & & \\
  \hline
  V2540~Oph & & & & & \\
  Atomic identification & Rest wavelength (\AA) & 4/12 & 5/19 & & \\
  \hline
  H$\beta$  & 4861 & & $-1360 \pm 150$ & & \\
  Fe~II      & 5018 & $-1540 \pm 100$ & $-1120 \pm 100$ & & \\
  Fe~II      & 5169 & $-1420 \pm 100$ & $-1160 \pm 100$ & & \\
  \hline

  & & & & & \\
  \hline
  V5113~Sgr & & & & & \\
  Atomic identification & Rest wavelength (\AA) & 9/21 & 9/27 & 10/8 & 10/24 \\
  \hline

  H$\beta$  & 4861 & & $-1180 \pm 100$ & & \\
  Fe~II      & 5018 & & $-1070 \pm 100$ & & \\
  Fe~II      & 5169 & $-1380 \pm 200$ & $-820 \pm 100$ & $-1290 \pm 100$ & \\
  Na~I       & 5896 & $-740 \pm 100$ & $-1250 \pm 100$ & $-1410 \pm 100$ & $-1800 \pm 100$ \\
  H$\alpha$  & 6563 & & $-1070 \pm 100$ & & \\
  \hline

  & & & & & \\
  \hline
  V4745~Sgr & & & & & \\
  Atomic identification & Rest wavelength (\AA) & 5/8 & 5/28 & & \\
  \hline
  H$\beta$  & 4861 & $-1650 \pm 100$ & $-2030 \pm 100$ & & \\
  Fe~II      & 5018 & $-1640 \pm 100$ & $-2480 \pm 150$ & & \\
  Fe~II      & 5169 & $-1740 \pm 200$ & & & \\
  Na~I       & 5896 & $-1600 \pm 100$ & & & \\
  H$\alpha$ & 6563 & $-1740 \pm 100$ & $-2230 \pm 100$ & & \\
  \hline

  & & & & & \\
  \hline
  V458~Vul & & & & & \\
  Atomic identification & Rest wavelength (\AA) & 8/9 & 8/10 & & \\
  \hline
  H$\beta$  & 4861 & $-1720 \pm 100$ & $-1930 \pm 100$ & & \\
  He~I      & 5015 & $-1480 \pm 100$ & & & \\
  He~I      & 5876 & $-1360 \pm 100$ & & & \\
  H$\alpha$ & 6563 & $-1920 \pm 100$ & $-1840 \pm 100$ & & \\
  He~I      & 6678 & $-1450 \pm 100$ & & & \\
\end{longtable}

\section{Discussion}
\subsection {Properties of the early rebrightenings of the light curve}
We obtained the spectra of 6 classical novae with several rebrightenings during the early phase. The light curve of V1186~Sco is presented in figure \ref{v11861}. The amplitude of the first peak (JD 2453195.9) measured from the minimum before the second peak in the visual magnitude is about 1.2 mag. The amplitude, between the second peak and the minimum after the second peak (JD 2453204.9-2453209.5), is approximately 1.0 mag. The last rebrightening reached its maximum on JD 2453213.1, only $\sim$17 days after the peak of the first maximum on JD 2453195.9.

\begin{figure}
  \begin{center}
    \FigureFile(80mm,50mm){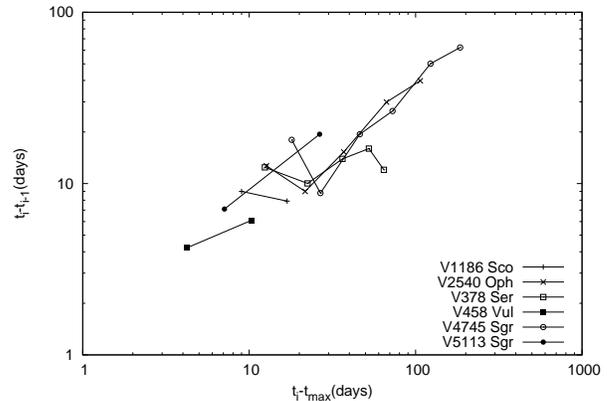}
  \end{center}
  \caption{Time interval between two successive maxima $t_i-t_{i-1}$ as a function of time from the initial optical maximum $t_i-t_{\rm max}$.}\label{interval}
\end{figure}

The light curve of V2540~Oph is presented in figure \ref{v25401}. The average peak-to-peak amplitude is approximately 1.5 mags and the duration of the rebrightening phase was about four months after the discovery. These rebrightenings show six or seven peaks and the time interval between successive maxima increases gradually (figure \ref{interval}). Pejcha (2009) also reported this trend of the interval elongation. The mean visual magnitude of V2540~Oph scarcely varies during the rebrightenings.

The light curve of V5113~Sgr is presented in figure \ref{v51131}. It shows three peaks during the early phase. The amplitude of the first rebrightening is approximately 2.2 mag (JD 2452901.8-2452903.9), and the average amplitude is approximately 1.8 mag. This light curve is different from the light curves of the other novae in this paper, in that the rise of the rebrightening is slower than the decline. The duration of the rebrightening phase is about one month.

The light curve of V4745~Sgr is presented in figure \ref{v47451}. It shows more than seven peaks. It seems similar to the light curve of V2540~Oph during the early phase (figure \ref{v25401}), while the mean visual magnitude slowly declines during the rebrightening phase in V4745 Sgr. The average amplitude is approximately 2.6 mag except the second mini outburst (JD 2452761.8) and the duration of the early rebrightening phase is about six months. The time interval between successive maxima increases gradually during the early phase. Cs$\acute{\rm a}$k et al.~(2005) reported this increasing nature of the recurrent time, which was also reported previously in GK Per and DK Lac by Bianchini et al.~(1992).

The light curve of V458~Vul is presented in figure \ref{v4581}. It exhibits three peaks with the average amplitude of about 1.8 mags during the early phase, and the rebrightenings with a smaller amplitude during the transition phase after JD 2454370. The duration of the rebrightening phase is approximately two weeks (JD 245322-245335). Note that this light curve sometimes shows small oscillations with an amplitude of smaller than 1 mag.

The light curve of V378~Ser is presented in figure \ref{v3781}. It exhibits many rebrightenings or oscillations during the early phase. The amplitude between the third maximum and the minimum after that maximum (JD 2453476.3-2453478.2) is about 1.7 mags and the average amplitude is approximately 1.5 mag.

V1186~Sco, V5113~Sgr, and V458~Vul are identified as fast novae, while V2540~Oph, V4745~Sgr and  V378~Ser as slow or moderately fast novae. A clear difference between the former and latter is the number of experienced rebrightenings, $\sim$3 in the fast novae, while $\sim$6 in the slow novae. The duration of the rebrightening phase is $\sim$10 days in the fast novae, while $\sim$100 days in the slow novae. 
The rebrightenings during the early decline have been also observed in the novae V1178 Sco (moderately fast nova), V4361 Sgr (slow nova) (\cite{Kato2001}), V2214 Oph (slow nova) (\cite{Lynch1989}), and V868 Cen (fast nova) (\cite{Williams2003}). The light curves of these novae trace this trend of the duration of the rebrightening phase.

As noted above, the time interval between successive maxima increases gradually (see \cite{Csak2005}; \cite{Pejcha2009}). Strope et al.~(2010), on the other hand, reported that it appears that the timing of the rebrightenings is random within the interval over which they occur. We plotted the successive recurrence times versus time from the first maximum in figure \ref{interval}. We can find that the time intervals of the rebrightenings of our 6 targets exhibit a similar systematic trend. We first confirmed that this trend is applicable to all types of novae: fast and slow, and Fe~II type and hybrid type. We find that this trend follows the equation:
\begin{equation}
 \log(t_i-t_{i-1}) = a \log(t_i-t_{max}),
\end{equation}
where $a$ is the slope. The least-squares fit to all data of 6 novae yields $a = 0.79 \pm 0.01$. Pejcha~(2009) derived $a$ of $0.88 \pm 0.04$ for DK Lac and $0.79 \pm 0.04$ for V4745 Sgr. The values are close to that of our result.

\subsection{Common character of the spectral evolutions}
In 6 novae, V1186~Sco, V2540~Oph, V5113~Sgr, V4745~Sgr, V458~Vul and V378~Ser, we found that some of the lines in the spectra at and around the brightness maximum show P-Cygni profiles, and they do not have the absorption components of the P-Cygni profile in the interval between successive maxima. These facts suggest the following scenario. First, the photosphere of the nova rapidly expands after the nova eruption. The temperature of the photosphere decreases as the photosphere expands. Next, when the radius of the optically thick envelope becomes maximum, the nova reaches its maximum brightness in the optical and the spectrum has stronger absorption lines. Although the optically thin envelope continue to expand, the photosphere then start to shrink. The continuum of the optical spectrum decays, as the photosphere shifts inside. Consequently, the absorption components disappear and then the emission lines become stronger due to decay of the optical continuum, and expansion of the optically thin gas envelope. After that, the optical continuum rises again, without growth of the emission lines, in some novae. This behavior proves reexpansion of the photosphere. 

Based on the observations of V4745 Sgr, Cs$\acute{\rm a}$k~et~al.~(2005) mentioned that the reappearance of the P-Cygni profile suggested mass ejection at the rebrightening. Pejcha~(2009) concluded that the rebrightening were likely caused by hydrogen-burning instabilities. As noted above, we proposed that rebrightening without growth of the emission lines, and reappearance of the absorption component of the P-Cygni profile indicate re-expansion of the photosphere, which is not inconsistent with the interpretations by Cs$\acute{\rm a}$k~et~al.~(2005) and Pejcha~(2009).

Table \ref{tab_pcygni} summarize the variability of the blue-shift velocity of the absorption component of the P-Cygni profiles in 5 novae except V378~Ser. We can not find a common trend of this variabilities. In V2540~Oph, all the blue-shift velocities of Fe~II lines decreased with time. In V5113~Sgr, the velocities of Fe~II~5169 do not exhibit a simple trend, while the velocities of Na~I increased progressively. In V4745~Sgr, all the velocities increased greatly.

\subsection{Comparison with the second rebrightening in V2362 Cyg}
There are several types of the light curves of novae, although no two novae show exactly the same light curves. In this paper, we presented the light curve of the 6 novae showing a few to several rebrightenings during the early phase. In this and next subsections, we will compare these novae to some different types of novae. 

V2362 Cyg, a slow nova, caused the rebrightening eight months after the initial maximum (see \cite{Kimeswenger2008}; \cite{Lynch2008}; \cite{Poggiani2009}). Munari et al.~(2008) presented the spectral evolution of V2362 Cyg. They reported that, at the second maximum, the underlying continuum was much hotter and the absorption lines, He~I lines (for example 5876 \AA), were present. This is inconsistent with our idea that reappearance of the P-Cygni profile is related to re-expansion of the photospere, since the photosphere expansion means decrease of the effective temperature. The mechanism of the rebrightenings observed in our targets is thought to be different than that in V2362 Cyg.

Similar rebrightenings long time after the maximum have been also observed in V1493 Aql and V2491 Cyg (see e.g., \cite{Bonifacio2000}; \cite{Kiyota2004}; \cite{Hachisu2009}). Hachisu \& Kato (2009) suggested that the mechanism of the rebrightening is the strong magnetic reconnection between the white dwarf and the companion star. Kato et al.~(2009) reported that the light curves of V1493 Aql and V2362 Cyg are composed of the power-law decline and exponential brightenings. They proposed that the rebrightening can be caused by a shock resulting from a secondary ejection and its breakout in the optically thick nova winds. 

\subsection{Comparison with the oscillation during the transition phase in V1494 Aql}
The light curves of some novae, like the fast nova V1494 Aql, show oscillations during the transition phase, at about 3.5 magnitude below the brightness maximum (see e.g., \cite{McLaughlin1943}; \cite{Iijima2003}). This oscillation seemingly mimics rebrightenings during the early phase. Iijima and Esenoglu (2003) presented the spectra during transition phase of V1494 Aql. They suggested that the origin of this oscillation was high velocity jets, because high velocity broad components appeared in the red and blue sides of Balmer lines during the transition phase. Such high velocity wings, however, do not appear in our observations. Moreover, the emission lines in V1494 Aql did not have a P-Cygni profile at the maximum brightness. We thus suggest that the origin of rebrightenings during the early phase is different from that of the oscillations during the transition phase.

\section{Conclusions}
The light curves of the novae V1186~Sco, V2540~Oph, V5113~Sgr, V4745~Sgr, V458~Vul and V378~Ser display several rebrightenings during the early phase. Some slow or moderately fast, but not fast novae, have been known to show rebrightenings having a large amplitude during the early phase, although it is not clear how slow novae accomplish these rebrightenings. We found that three fast nova V1186~Sco, V458~Vul, and V5113~Sgr caused such rebrightenings. Moreover, the recurrence time of the rebrightening increase with a power law of the time from the first maximum (the index of 0.79(1)). This trend is applicable to all types of novae: fast and slow, and Fe~II type and hybrid type.

Our low-resolution optical spectra revealed spectroscopic changes during the early rebrightenings. The spectra at the first brightness maximum of these rebrightenings show many emission lines with P-Cygni profiles. We found that the absorption component of these P-Cygni profiles once disappear after the brightness maximum, and then reappear at the next brightness maximum, in some lines. We suggest that the rebrightening is due to enhancement of the continuum, and not due to growth of the flux of the emission lines. These means re-expansion of the photosphere at rebrightenings.

In addition, We compared 6 novae in our observations with other novae showing rebrightenigs/oscillations during the transition phase. We are unable to find the common features between our spectra and the spectra at the second maximum in V2362 Cyg, and the spectra during the transition phase in V1494 Aql. It suggests that the physical origin of the rebrightenings in our 6 targets in the early phase is different from that in V2362 Cyg and V1494 Aql.

\bigskip
The authors are thankful for observers who have posted their precious data to VSNET and AAVSO. We used the data which were obtained by the ASAS-3 system and made publicly available. Thanks are also to A. Imada, T. Ohshima and K. Imamura for their valuable comments. This work was supported by the Grant-in-Aid for the Global COE Program "The Next Generation of Physics, Spun from Universality and Emergence" from the Ministry of Education, Culture, Sports, Science and Technology (MEXT) of Japan. The authors are indebted to the anonymous referee for her/his useful comments.


\begin{thebibliography}{}

\bibitem[Ak et al.~(2005)]{Ak2005}
  Ak, Tansel, et al.~2005, PASA, 22, 298
\bibitem[Ashock \& Banerjee~(2003)]{Ashock2003}
  Ashock \& Banjerjee~2003, \iaucirc, 8132
\bibitem[Arai et al.~(2009)]{Arai2009}
  Arai, A., et al.~2009, The Eighth Pacific Rim Conference on Stellar Astrophysics: A Tribute to Kam-Ching Leung ASP Conference Series, 404, 90
\bibitem[Bianchini et al.~(1992)]{Bianchini1992}
  Bianchini, A., et al.~1992, \aap, 257, 599
\bibitem[Bonifacio et al.~(2000)]{Bonifacio2000}
  Bonifacio, P., et al.~2000, \aap, 356, 53
\bibitem[Brown et al.~(2003a)]{Brown2003a}
  Brown, J. et al.~2003a, \iaucirc, 8123
\bibitem[Brown et al.~(2003b)]{Brown2003b}
  Brown, N. J. et al.~2003b, \iaucirc, 8204
\bibitem[Buil \& Fujii~(2007)]{Buil2007}
  Buil, C. \& Fujii, M.~2007, \iaucirc, 8862
\bibitem[Burlak \& Henden~(2008)]{Burlak2008}
  Burlak, M.A.~\& Henden, A.A.~2008, Astronomy Letters, 34, 4, 241
\bibitem[Cs$\acute{\rm a}$k et al.~2005]{Csak2005}
  Cs$\acute{\rm a}$k, B., et al.~2005, \aap, 429, 599
\bibitem[Dobrotka et al.~(2006)]{Doborotka2006}
  Dobrotka, A., et al.~2006, \mnras, 371, 459
\bibitem[Duerbeck~(1981)]{Duerbeck1981}
  Duerbeck, H. W.~1981, PASP, 93, 165
\bibitem[Ederoclite et al.~(2005)]{Ederoclite2005}
  Ederoclite, A., et al.~2005, \iaucirc, 8506
\bibitem[Hachisu \& Kato~(2009)]{Hachisu2009}
  Hachisu, I. \& Kato, M.~2009, \apj, 694, 103
\bibitem[Hellier~(2001)]{Hellier2001}
  Hellier, C.~2001, Cataclysmic Variable Stars: how and why they vary
  (Berlin: Springer-Verlag)
\bibitem[ Iijima \& Esenoglu~(2003)]{Iijima2003}
  Iijima, T. \& Esenoglu, H. H.~2003, \aap, 404, 997
\bibitem[Kamath et al.~(2005)]{Kamath2005}
  Kamath, U. S., et al.~2005, \mnras, 361, 1165
\bibitem[Kato \& Fujii~(2001)]{Kato2001}
  Kato, Taichi \& Mitsugu, Fujii~2001, IBVS, 5150
\bibitem[Kato et al.~(2002)]{Kato2002}
  Kato, T., et al.~2002, IBVS, 5309
\bibitem[Kato et al.~(2004)]{Kato2004}
  Kato, T., et al.~2004, \pasj, 56, S1
\bibitem[Kato et al.~(2009)]{Kato2009}
  Kato, T., et al.~2009, Var. Star. Bull., 49, 1
\bibitem[Kimeswenger et al.~(2008)]{Kimeswenger2008}
  Kimeswenger, S., et al.~2008, \aap, 479, 51
\bibitem[Kiyota et al.~(2004)]{Kiyota2004}
  Kiyota, S., et al.~2004, \pasj, 56, S193
\bibitem[Lynch et al.~(1989)]{Lynch1989}
  Lynch, David K., et al.~1989, \aj, 98, 1682
\bibitem[Lynch et al.~(2008)]{Lynch2008}
  Lynch, David K., et al.~2008, \aj, 136, 1815
\bibitem[McLaughlin et al.~(1943)]{McLaughlin1943}
  McLaughlin, et al.~1943, Publications of the Observatory of the University of Michigan, 8, 149
\bibitem[Munari et al.~(2008)]{Munari2008}
  Munari, U., et al.~2008, \aap, 492, 145
\bibitem[Nakamura et al.~(2002)]{Nakamura2002}
  Nakamura, Y., et al.~2002, \iaucirc, 7808
\bibitem[Nakano et al.~(2007)]{Nakano2007}
  Nakano, S., et al.~2007, \iaucirc, 8861
\bibitem[Payne-Gaposchkin~(1964)]{Payne1964}
  Payne-Gaposchkin, C.~1964, The Galactic Novae (Dover, New York)
\bibitem[Pejcha~(2009)]{Pejcha2009}
  Pejcha, Ond$\check{\rm r}$ej~2009, \apj, 701, 119
\bibitem[Poggiani~(2008)]{Poggiani2008}
  Poggiani, R.~2008, \apss, 315, 79
\bibitem[Poggiani~(2009)]{Poggiani2009}
  Poggiani, R.~2009, New Astronomy, 14, 4
\bibitem[Pojmanski 2002]{Pojmanski2002}
  Pojmanski, G.~2002, Acta Astronomica, 52, 397
\bibitem[Pojmanski et al.~(2004)]{Pojmanski2004}
  Pojmanski, G., et al.~2004, \iaucirc, 8369
\bibitem[Pojmanski et al.~(2005)]{Pojmanski2005}
  Pojmanski, G., et al.~2005, \iaucirc, 8505
\bibitem[Retter et al.~(2002)]{Retter2002}
  Retter, A. et al.~2002, \iaucirc. 7809
\bibitem[Rodr$\acute{\rm \i}$guez-Gil et al.~(2010)]{Rodriguez2010}
  Rodr$\acute{\rm \i}$guez-Gil, P., et al.~2010, MNRAS, 407, 1, L21
\bibitem[Ruch et al.~(2003)]{Ruch2003}
  Ruch, G., et al.~2003, \iaucirc, 8204, 2
\bibitem[Rudy et al.~(2004)]{Rudy2004}
  Rudy, R. J., et al.~2004, \iaucirc, 8370, 2
\bibitem[Schwarz et al.~(2007)]{Schwarz2007}
  Schwarz, G. J., et al.~2007, \aj, 134, 516
\bibitem[Seki et al.~(2002)]{Seki2002}
  Seki, T., et al.~2002, \iaucirc, 7809
\bibitem[Smith \& Dhillon~(1998)]{Smith1998}
  Smith, D. A.~\& Dhillon, V. S.~1998, MNRAS, 301, 3, 767
\bibitem[Spencer~1931]{Spencer1931}
  Spencer J. H. Cape Obs. Ann. 10, No. 9.
\bibitem[Starrfield et al.~(1976)]{Starrfield1976}
  Starrfield, S., et al.~1976, IAUS, 73, 155
\bibitem[Strope et al.~(2010)]{Strope2010}
  Strope, R. J., et al.~2010, \apj, in press 
\bibitem[Tarasova~(2007)]{Tarasova2007}
  Tarasova, T. N.~2007, IBVS, 5807, 1
\bibitem[Vogt (1928)]{Vogt1928}
  Vogt, H. 1928, Astronomische Nachrichten, 232, 269
\bibitem[Williams et al.~(1991)]{Williams1991}
  Williams, R. E., et al.~1991, \aj, 376, 721
\bibitem[Williams~(1992)]{Williams1992}
  Williams, R. E.~1992, \aj, 104, 725
\bibitem[Williams et al.~(2003)]{Williams2003}
  Williams, R. E., et al.~2003, THE JOURNAL OF ASTRONOMICAL DATA, 9

\end{thebibliography}
\end{document}